\documentstyle[11pt,fewbody2005,twoside,epsfig]{article}

\begin{document}

\title{The Slingshot Revisited}

\author{Sverre Aarseth}
\affil{Institute of Astronomy, University of Cambridge, England}

\begin{abstract}

Some 30 years ago, the paper by Saslaw, Valtonen and Aarseth
introduced the term ``gravitational slingshot'' into the literature. 
This concept was invoked to explain the properties of double radio sources
by means of ejection from a strongly bound triple system, where a binary
was also ejected by recoil.
Since then several investigations have illustrated the general behaviour of
such triple interactions in different types of $N$-body systems, although
not much detailed analysis is available.

In this celebratory contribution, we present some new data from standard
star cluster modelling containing primordial binaries and triples, as well
as results from a binary black hole simulation with two massive members.
In the star cluster case, the process of mass loss from evolving stars,
together with general mass segregation, promotes favourable interactions 
involving compact subsystems of binaries and triples in the central region.
Three-body interactions often lead to energetic ejections, with one or
more of the components attaining relatively large terminal velocities.
Hence the presence of primordial binaries in both open clusters and the
richer globulars inevitably produces some high escape velocities which can
be observed in principle.

The second type of stellar system to be discussed is based on the scenario
of two approaching galactic cores with density cusps, each having a
massive black hole.
After the subsystems become well mixed, the two massive components soon
form a hard binary which gains energy by ejecting other members.
Such a massive binary has a large cross section and can be very effective
in depleting bound stars from the core.
Again high-velocity escapers are produced, with their terminal speeds
related to the shrinking binary size.
Although the large mass ratio prevents escape by recoil here, the Brownian
motions of the binary exceed the predicted values significantly.
This wandering is due to the small restoring force near the centre and has
implications for the so-called loss cone depletion.

\end{abstract}

\section{Introduction}

The original motivation for the slingshot paper (Saslaw, Valtonen \&
Aarseth 1974) was to explain the properties of double radio sources by the
ejection of massive bodies from galactic nuclei.
Scaling down the masses, we can find many interesting applications in star
clusters.
It is now thought that both open clusters and globulars contain an appreciable
fraction of primordial binaries.
The subsequent dynamical cluster evolution therefore gives rise to a large
number of interactions between single stars and binaries, as well as between
the binaries themselves.
If a hard binary is involved in a strong triple encounter, the final velocity
of the escaper may readily exceed the typical velocity of single stars.
Likewise, the binary may escape as one object by the recoil effect, and the
same goes for the interaction of two binaries.
On the other hand, a somewhat larger impact parameter may lead to the widest
binary being disrupted, with one component captured into a stable or
long-lived orbit around the other binary.
Such dormant triple systems may become dynamically active at a later stage,
following increase of the outer eccentricity or mass loss by the inner binary.
This second type of slingshot event is also of considerable interest because
a population of stable triples may be present in clusters containing
primordial binaries (Aarseth 2004).

In the following, we describe some results of realistic star cluster
simulations with a small primordial population of stable triples in addition
to a larger fraction of binaries.
Such systems have not been studied up to now and a series of models were
investigated in order to carry out comprehensive code testing.
This type of initial conditions seems justified because of the significant
proportion of triples and higher multiples observed in the field, while their
detection inside clusters is hampered by selection effects and the longer
periods of the outer orbit.
One section is also devoted to describing some results from a recent numerical
study of a massive binary formed by the merging of two separate stellar
systems (Aarseth 2003b).

\section{Binary dynamics}

We begin by summarizing some appropriate quantities for the study of binaries
in clusters.
The concept of a hard binary (Heggie 1975) is of fundamental importance.
An approximate expression for the semi-major axis  is obtained by comparing
the binding energy, $E_b$, in scaled units with the mean kinetic energy,
\begin{equation}
  \frac{m_1 m_2}{2 a_h} = \frac{1}{2} {\bar m} V^2 \,,
\end{equation}
with ${\bar m} = 1/N$, which gives $a_h = 2 r_h/N$ for an equal-mass system
of half-mass radius $r_h$ in virial equilibrium.
It is also known theoretically that the hardening rate is constant, with the
most significant changes occurring in cascades (Heggie 1975).
This allows us to illustrate the slingshot mechanism by combining two local
conservation relations,
\begin{equation}
 E_b + \frac{1}{2} \mu v_3^2 \simeq {\rm const} \,,~~~~~~~
 \sum m_i {\bf v}_i \simeq {\rm const} \,,
\end{equation}
with $\mu$ the reduced mass of $m_3$ and $v_3$ the corresponding relative
velocity.

If the binary hardening proceeds without an intervening collision, the change
in binding energy may be sufficiently large for even the binary to recoil
out of the cluster.
Hence the sudden removal of one or more stars has an equivalent effect to
that of rapid mass loss by supernovae explosion.
Since such events invariably occur near the centre, the process of core
collapse expected on theoretical grounds is much alleviated.
In view of the $m^2$ dependence of $E_b$, the importance of a mass spectrum
should also be emphasized, especially since the time-scale for mass
segregation goes as $1/m$.
From theoretical considerations, the typical gain in binding energy in a
strong interaction is about $40 \,\%$ for equal masses.
Hence the slingshot condition connects the relative escape velocity of a
single particle with the semi-major axis by the approximate expression
\begin{equation}
   v_f \simeq \left ( \frac {m_1 + m_2}{2 a} \right )^{1/2} \,.
\end{equation}
Because of the eccentricity effect (e.g. encounters near small pericentre),
the outgoing velocity may on rare occasions exceed this value by a
considerable amount.
We therefore speak about super-fast escape from clusters containing
super-hard binaries, with $E_b$ approaching the total energy in some
clusters.

Long-lived hierarchical systems also play an important role for cluster
dynamics.
To facilitate the numerical treatment, we define a triple to be stable if
the outer pericentre exceeds a critical value
\begin{equation}
{R_p^{crit}} ~=~ C\left[(1+q_{out})\frac{(1+e_{out})}
{(1-e_{out})^{1/2}}\right]^{2/5}a_{in} \,,
\end{equation}
where $q_{out} = m_3/(m_1 + m_2)$, $a_{in}$ is the inner semi-major axis
and $C \simeq 2.8$ is a numerical fitting constant appropriate for planar
motion (Mardling \& Aarseth 1999).
This criterion has been tested extensively in the absence of external
perturbations and appears to be surprisingly robust.
At present a small empirical correction (up to $30 \%$) is made for the
inclination effect and the weak dependence on the inner mass ratio and
eccentricity is not included.
Systems satisfying this inequality are treated in the two-body
approximation, with the inner semi-major axis assumed to be constant,
which neglects small periodic oscillations.
Consequently, any subsequent increase of the outer eccentricity, $e_{out}$,
may invalidate the stability condition, in which case the inner binary is
re-initialized as a two-body solution.

The numerical integration of hard binaries employs the Kustaanheimo--Stiefel
(1965) two-body regularization, while strong interactions of 3--5 particles
are treated by chain regularization (Mikkola \& Aarseth 1993).
Finally, the fourth-order Hermite integration scheme is used to advance the
solutions of the single particles (Aarseth 2003a).

\section{Cluster simulations}

In the present investigation of the slingshot effect we study a number of
similar cluster parameters.
All the models have initial populations with memberships
$N_s = 3500, N_b = 500, N_t = 100$, where $N_s, N_b, N_t$ denote single
stars, binaries and triples, respectively.
The single stars are distributed in an equilibrium Plummer model, with
the masses sampled from a Salpeter IMF in the range $[0.2 - 10]~M_\odot$.
The binary distribution is taken to be flat for $log (a)$ within
$[20 - 200]$~au.
A fraction of the binaries are converted to triples satisfying the stability
condition (4), with the inner semi-major axes first chosen at random in
$[8 - 80]$~au and reduced by factors of 2 if necessary.
Since a conventional tidal field for open clusters is applied, we adopt a
virial scale factor of 2~pc which corresponds to a half-mass radius
$r_h = 1.6$~pc.
For a realistic star cluster simulation, synthetic stellar evolution with
mass loss is included (Aarseth 2003a).
The prescription for supernovae explosions involves assigning a velocity
kick (limited to about 25 km/s here) to the small number of neutron stars
with progenitor masses above $8~M_{\odot}$.
However, these events usually occur during the first 50~Myr while most
slingshot velocities are found at later times.

\begin{figure} [h]
\begin{center}
\epsfig{file=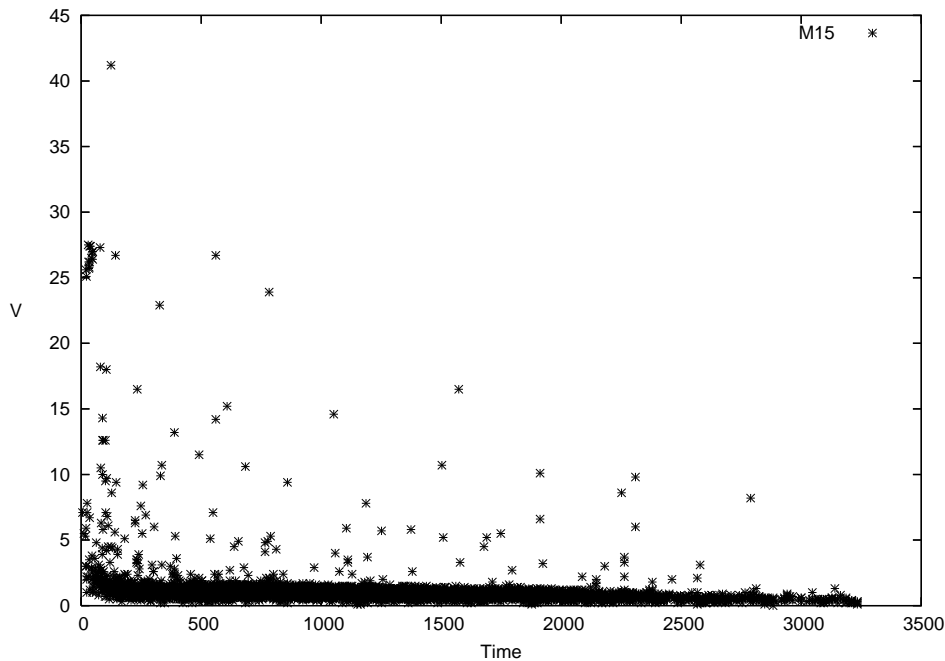}
\caption{Escape velocity in km/s vs time in Myr, model \#15.} \label{fig-1}
\end{center}
\end{figure}

A typical plot of the terminal escape velocities is shown in Fig.~1 for
the whole cluster life-time of 3.3~Gyr.
We may define a dimensionless value of energetic escape by
$\alpha = v_{\infty}^2/V^2$, where $V$ is the rms velocity.
The latter decreases from about 2.0~km/s initially to below 1.0~km/s when
half the stars have escaped at 1.1~Gyr.
Consequently, an appreciable fraction of the escapers are ejected with
velocities that may be said to be super-fast.
Comparable graphs for similar clusters without primordial binaries
show very little evidence for such high velocities.
Likewise, there are considerably fewer large escape velocities when
only binaries are included in the models, with typical maximum values
near 10~km/s.

The question of how the relative memberships of binaries and triples behave
with time is of fundamental interest and there has been various claims in
the literature.
Over 20 models have been examined and the similarities are remarkable.
The three populations are plotted in Fig.~2 as a function of time until
complete disruption.
Contrary to expectations, the single stars show clear evidence for a
preferential depletion.
It is particularly interesting that the triples are quite robust in spite
of being centrally concentrated (Aarseth 2004).
Moreover, the presence of even a small binary fraction has the effect of
preventing core collapse which in any case is much diluted if mass loss
from evolving stars is taken into account.

\begin{figure} [h]
\begin{center}
\epsfig{file=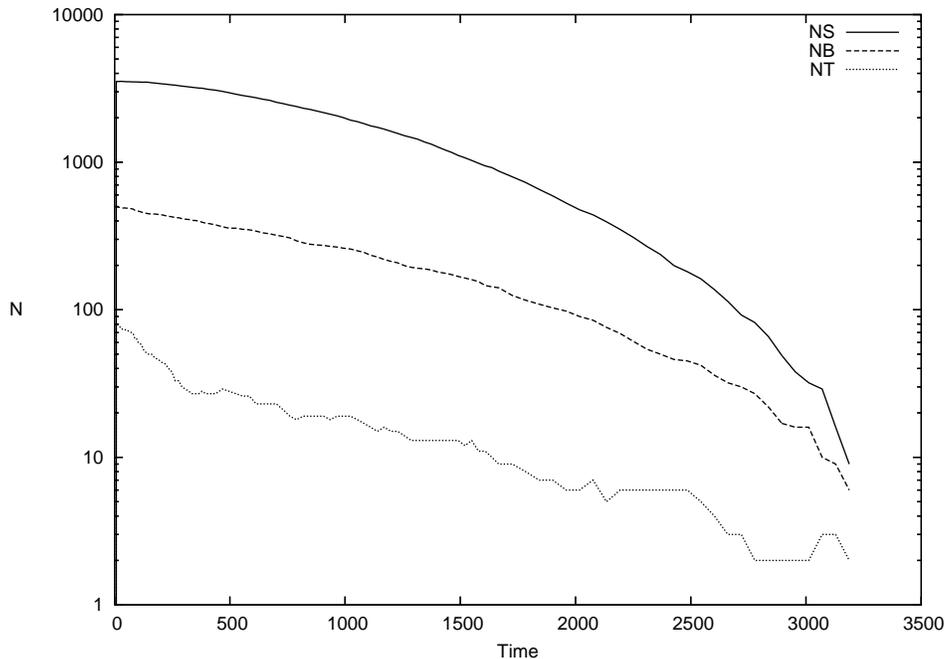}
\caption{Single stars, binaries and triples vs time in Myr, model \#15.}
 \label{fig-2}
\end{center}
\end{figure}

\section{Slingshot events}

The large velocities featured in Fig.~1 are undoubtedly associated with the
slingshot mechanism.
It is instructive to examine some examples in detail to illustrate the
recoil effect.
For this purpose we describe three different few-body interactions which
resulted in relatively large ejection velocities.

A classical slingshot event occurs in a fly-by, where the incoming body
gains kinetic energy in one passage and the binary is ejected in the
opposite direction.
In the first case, a binary with period $P = 190$~d and equal masses of
$3.0~M_{\odot}$ increased its energy to $P = 33$~d and characteristic high
eccentricity, $e = 0.88$.
This resulted in respective terminal velocities of 25 and 65~km/s.
Since these escapers are removed at similar times, they may be associated
with a common origin.
Including the third body of mass $m_3 = 2.4~M_{\odot}$ in the momentum
check yields $\sum m_i v_i / \sum m_i \simeq 0.6$~km/s, which is well
within uncertainties when assuming zero centre of mass (c.m.) velocity.
Note that such high-velocity ejections are essentially radial, hence the
scalar velocity $v_i$ (with opposite sign) is sufficient here.

In the alternative case of exchange, the initial period was 131~d, with
masses $2.8, 2.8, 6.1~M_{\odot}$.
Following exchange by the heavy intruder, the period was reduced to 71~d
and moderate final eccentricity $e = 0.51$.
Applying the momentum check gives a c.m. value of 1.4~km/s from the
corresponding final velocities 18 and 62~km/s.
Regarding the energetics, we note that according to Eq.~(3), a typical
large escape velocity of 10~km/s may be produced in an interaction
with a $2 M_{\odot}$ binary of size 10~au which is consistent with the
adopted period distribution.

A third energetic example involved the impact of two formerly stable
triple systems.
The long outer periods, $P_{out} = 20\,000$ and $60\,000$~d and low-mass
components, $m_3 = 0.2~M_{\odot}$, resulted in prompt ionization.
The subsequent strong interaction of the inner binaries with respective
periods 20 and 1250~d produced one final binary of period $P = 14$~d and
an unusually low eccentricity, $e = 0.2$.
Given the initial and final masses $2.0, 2.0, 3.8, 3.8~M_{\odot}$ and
$5.8, 3.8, 2.0~M_{\odot}$ shows that exchange has occurred.
The component velocities of the three objects were all substantial at
21, 30 and 57~km/s, respectively.
In this case, however, momentum conservation does not apply in its simple
form and the velocity vectors are not available for the consistency check.

\section{Massive binary evolution}

Stellar systems containing a massive binary are of considerable topical
interest.
Such simulations were already made a long time ago (Aarseth 1972), when
it was demonstrated that a binary containing massive components
($25 {\bar m}$) absorbed 90\,\% of the total energy after only 50 crossing
times.
Moreover, other models with $N = 250$ and smaller mass ratios acquired
50\,\% of the total energy in less than 20 crossing times.
It was also noted that this evolution gave rise to large escape velocities,
with occasional values up to $v_{\infty} \simeq 5V$.

This problem was revisited in another context recently (Aarseth 2003b).
Here we adopt a cosmological scenario with two interacting dwarf galaxies,
each containing one massive central object assumed to be a black hole.
As expected, merging leads to the rapid formation of a massive binary
which continues to shrink with time.
Two equilibrium systems with $N_0 = 120\,000$ equal-mass particles each
were placed in an initial orbit of eccentricity $e = 0.8$ and separation
$8 r_h$.
The black holes were at rest locally and assigned a mass
$m_{BH} = (2 N_0)^{1/2} {\bar m}$.

The integration of the single particles was performed on the GRAPE-6
special-purpose computer while the binary was treated by an accurate
time-transformed leapfrog method which also includes the three first
terms of the post-Newtonian approximation (Mikkola \& Aarseth 2002).
Because of the computational cost, the GR terms are only included if
the associated time-scales are sufficiently short.
A switch from two-body regularization is made when the binary becomes
super-hard near $a_{BH} \simeq 10^{-4} r_h$.
The special treatment includes direct interactions with other members
inside $25 a_{BH}$ as well as perturbers at twice this distance.

\begin{figure} [h]
\begin{center}
\epsfig{file=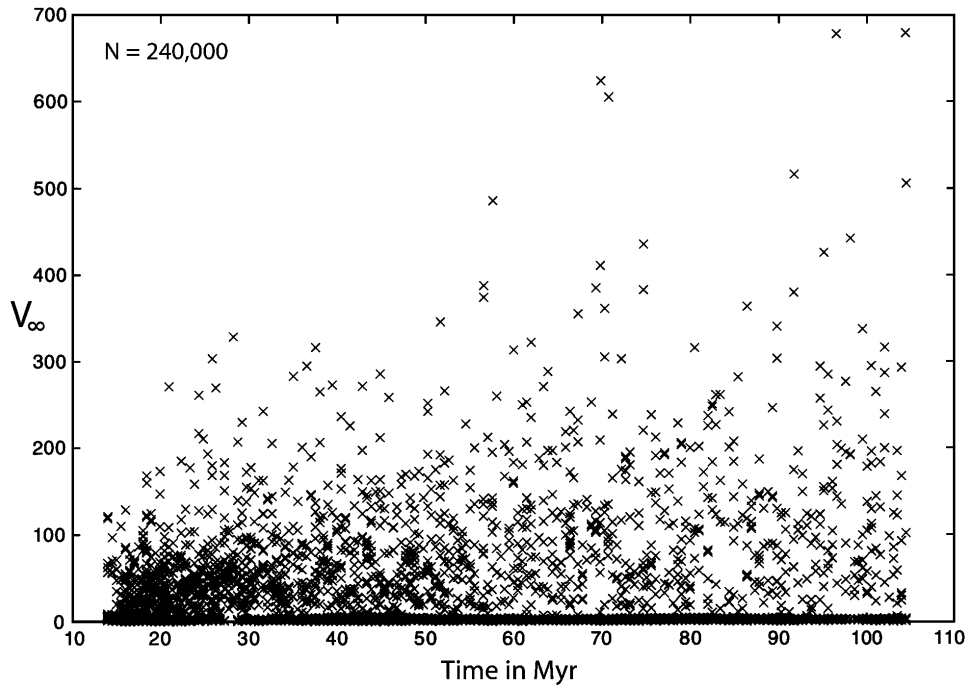,width=130mm}
\caption{Escape velocity in km/s vs time in Myr.}
 \label{fig-3}
\end{center}
\end{figure}

The subsequent binary shrinkage is connected with the ejection of
high-velocity particles by the slingshot mechanism.
Adopting the scaling $r_h = 4$~pc and ${\bar m} = 1~M_{\odot}$ yields
a velocity scale factor $V^{\ast} \simeq 16$~km/s for the rms velocity.
Figure~3 shows a plot of the terminal escape velocity.
The number of super-fast escapers is quite large, with maximum values
up to 600~km/s.
With an effective mass ratio of about 1000, the binary recoil is not
sufficient for escape to occur.
However, the typical c.m. velocity is significantly above the
equipartition condition.
Hence in this system, core wondering is effective and leads to
loss-cone replenishment.
A qualitative reason for this behaviour is due to the small gradient
of the potential at the centre of spherical systems, so that the
restoring force does not constrain the c.m. recoil.

Shrinkage by dynamical effects is not sufficient to induce black
hole coalescence for the present parameters.
However, GR effects may become relevant if substantial eccentricity
growth takes place.
In the present model, the eccentricity exceeded 0.995 in the later
stages (with $a_{BH} \simeq 2.4 \times 10^{-5}$) when the GR
terms were activated based on time-scale considerations.
At the end of the calculation, $a_{BH} \simeq 1 \times 10^{-5} r_h$
which necessitated a large number of Keplerian periods.
The calculation was terminated when coalescence occurred inside
three Schwarzschild radii after some 400 $N$-body time units
or 100~Myr.

The form of Eq.~(3) also applies to the case of large mass ratios
(like comets) when a velocity criterion rather than the energy gain
is appropriate.
Substituting an extreme value, $a_{BH} = 1 \times 10^{-5}$
(with $r_h \simeq 1$ in scaled units) yields $v_f \simeq 20$ which
corresponds to about 300~km/s.
We may therefore conclude that the most energetic escapers have been
produced with the massive binary well inside the semi-major axis,
especially since the binary size was a factor of 10 larger during
the early stage.
The slingshot mechanism still needs to be examined in detail.
Thus it appears that there is a preferential loss of angular
momentum during the strongest interactions which tend to be
associated with prograde motion.
Understanding this behaviour presents a challenge for theory.

\section{Conclusions}

This investigation of the slingshot mechanism has demonstrated that
surprisingly large velocities can be generated in self-consistent
stellar systems containing binaries with short periods.
For conventional open clusters, this feature is enhanced by including
primordial triples.
Although the corresponding cross section is very small, gravitational
focusing plays an important role.
A plot of escape velocity versus mass shows that high velocities are
not limited to small masses.
This behaviour is connected with mass segregation which increases the
central concentration of massive particles.

The presence of one heavy binary also leads to interesting developments.
We emphasize the short time-scale of formation when starting with two
separate systems in bound orbit.
For the modelling of massive black holes it is necessary to increase the
membership as far as possible since scaling to realistic conditions is
rather uncertain.
The model discussed here would in fact be appropriate for a star cluster
with two black holes of mass $500~M_{\odot}$ and some $10^5$ single stars.
In such a system, the cross section is relatively large which gives rise
to favourable interactions for producing high escape velocities.

The question of observational implications should also be considered.
In the past, high-velocity stars were usually excluded as non-members.
However, a careful search for candidates on the opposite side might
yield a meaningful coincidence.
A systematic survey of slingshot events in realistic star cluster models
is much overdue and highly desirable.
Finally, we mention a different type of slingshot involving a binary
in close encounter with a massive black hole at the Galactic centre,
where one disrupted component may be ejected with an unusually high
velocity and the other captured (Hills 1988).

\end{document}